\documentclass{article}

\usepackage{arxiv}
\usepackage{xcolor}
\usepackage[utf8]{inputenc}
\usepackage[T1]{fontenc}
\usepackage{hyperref}
\usepackage{url}
\usepackage{booktabs}
\usepackage{amsfonts}
\usepackage{amsmath}
\usepackage{nicefrac}
\usepackage{microtype}
\usepackage{graphicx}
\usepackage{natbib}
\usepackage{doi}
\usepackage{amssymb}
\usepackage{array}

\renewcommand{\headeright}{Preprint}
\renewcommand{\undertitle}{Preprint -- \today}
\renewcommand{\shorttitle}{The Metric Fossil}

\newcounter{snapshot}

\title{The Metric Fossil: Emergent Spacetime from Asymmetric Projection}

\author{
  Jonathon Sendall\\
  OU Philosophy Department\\
  \texttt{jonathon.sendall@ou.ac.uk}\\[0.3em]
  \href{https://jonathonsendall162367.substack.com/p/1db2c5bd-f369-4076-9279-2b55f035f537}{\textcolor{blue}{Substack Article}}\\[0.5em]
}

\hypersetup{
pdftitle={The Metric Fossil: Emergent Spacetime from Asymmetric Projection},
pdfsubject={physics.gen-ph, gr-qc, quant-ph},
pdfauthor={Jonathon Sendall},
pdfkeywords={Emergent Spacetime, Quantum Gravity, Topology, Projection Lag, Dark Matter, Gate Algebra},
}

\begin{document}
\maketitle

\begin{abstract}
This paper develops a conditional framework for understanding the emergence of measurable physical structure from a pre-metric domain. Contemporary physics provides powerful and precise descriptions of relations among already-defined observables, yet offers comparatively little on the prior question of how observability, separability, and metric structure themselves arise. I propose that if three-dimensional spacetime is the result of an asymmetric projection from a non-orientable pre-geometric regime grounded in a minimal invariant, then a determinate and internally constrained set of consequences follows. These include: time reinterpreted as projection asymmetry rather than as a dimension or entropy gradient; matter as stabilised residue of projection rather than ontological primitive; quantum correlation as pre-separable unity dissolved by non-orientable topology; black holes as regimes of projection saturation rather than information sinks; dark matter as structured lag in the projection process rather than undetected particle species; and gravity as metric tension at sites of high projection density. The framework does not claim empirical confirmation. Its claim is that the proposal is internally coherent, structurally constrained, capable of generating non-trivial research directions, and that several phenomena currently treated as anomalous or paradoxical become expected consequences of the architecture rather than problems requiring additional postulates. An annex presents candidate formal objects and identifies research obligations for each consequence.
\end{abstract}


\section*{Methodological Preface: The Conditional Register}
This paper is written in what I shall call the \textit{conditional register}. It does not assert that the framework described is empirically established, nor that it constitutes a complete physical theory. It asserts something more modest and, I will argue, more philosophically interesting: that if a particular structural hypothesis is granted, then a determinate set of consequences follows, several of which dissolve persistent anomalies in physical theory without introducing additional entities.

The conditional register is not a retreat from rigour. It is a precise epistemic position. The paper's claim is that the proposal is:
\begin{enumerate}
    \item \textbf{Internally coherent}: its components do not contradict one another.
    \item \textbf{Structurally constrained}: the consequences are not arbitrarily stipulated but follow from the architecture.
    \item \textbf{Explanatorily non-trivial}: it addresses phenomena that resist standard treatments.
    \item \textbf{Formally generative}: it points toward specific mathematical obligations rather than closing them.
\end{enumerate}

This approach has distinguished precedent in foundations of physics. Much of the early work on the holographic principle, on quantum gravity's pre-geometric regimes, and on emergent spacetime operated in precisely this register: identifying structural possibilities with internal coherence before empirical traction was available \citep{jacobson1995, verlinde2011, padmanabhan2010}.

The proposal is not only conceptually organised. It already admits a candidate algebraic realisation. The projection $\mathcal{P}$ can be given more than metaphorical status by modelling it as a gate map: a normal, unital, completely positive, idempotent conditional expectation onto an admissible subalgebra, subject to covariance and no-signalling constraints. In this formalisation, the ``fossil record'' of $M^3$ corresponds to the accessible algebra on which all physical predictions are evaluated; the pre-metric domain $\mathcal{C}$ corresponds to the full algebra from which the gate compresses. The coherence witness $W(\rho,F) = \|F\rho(I-F)\|_1$ then measures cross-boundary coherence of a state relative to the gate, and the record fidelity gap $\Delta_T(\rho_F, R)$ measures how much expectation values change when degrees of freedom removed by projection are symmetrised away. Crucially, gate and readout mismatch produces structured rather than arbitrary distortion, and this distortion is quantitatively constrained by the commutator norm $\|[F,E]\|$, rather than freely varying. The formal role of this constraint in bounding projection residue is specified in Annex A.3b.

The paper proceeds as follows. Section 1 identifies the explanatory gap that motivates the framework. Sections 2 through 4 develop the three-part structural schema: the Invariant, the Closure, and the Projection. Sections 5 through 11 examine the consequences of the projection hypothesis if granted. Section 12 states the open problem that constitutes the genuine frontier of the framework. An annex presents candidate formal objects and research obligations.

\section{The Explanatory Gap}
Modern physics is extraordinarily successful at describing relations among observables within spacetime. Quantum mechanics formalises correlations, probabilities, and energy exchange between defined entities. General relativity describes the geometric structure of spacetime itself with remarkable precision. Quantum field theory synthesises both into a framework of extraordinary predictive power.

Yet both frameworks, and their various extensions, share a structural presupposition that is rarely made explicit: they begin with observables already defined, entities already separable, and a metric already in place. What they describe, with great precision, is what happens \textit{among things that already exist} and can be distinguished from one another.

The prior question (how do observables become observable, how does separability arise, how does metric structure emerge from whatever precedes it) is addressed patchwork fashion at best. The holographic principle gestures at a relationship between boundary and bulk \citep{susskind1995, thooft1993}. Loop quantum gravity posits a discrete pre-geometric structure. String theory invokes extra dimensions whose compactification yields observed physics. Emergent gravity programmes treat spacetime geometry as a thermodynamic phenomenon \citep{jacobson1995}.

Each of these addresses a facet of the problem. None offers a systematic account of the passage from a pre-metric regime to the metricated world of observable physics. The question is not merely technical. It bears on the foundations of physics at the deepest level: what kind of thing is space, what kind of thing is time, and what is the status of the entities that appear to inhabit both?

The present paper proposes a conditional answer. If spacetime, matter, and time are all consequences of a single asymmetric projection from a non-orientable pre-geometric regime, then each of these questions receives a unified structural response. The proposal does not eliminate the need for further work. It organises that work around a coherent architecture.

\section{The Projection Hypothesis}
The central hypothesis of this paper is as follows:

\begin{quote}
\textit{If three-dimensional metric structure arises through an asymmetric projection from a non-orientable pre-metric regime grounded in a minimal invariant, then the emergence of spacetime, matter, time, and the apparent multiplicity of physical entities can be understood as consequences of that projection.}
\end{quote}

Each element of this hypothesis requires unpacking:

\textbf{Asymmetric projection} means a mapping in which the structure of the pre-image domain is not preserved symmetrically in the image. Multiple pre-image configurations map to single stabilised outcomes. The mapping is surjective but not injective: information is resolved rather than transferred. This asymmetry is what gives the projection its directional character, and it is this directionality that we identify, in Section 5, with time.

\textbf{Non-orientable} means that the pre-metric regime does not possess a consistent notion of global orientation. On a non-orientable surface, there is no stable distinction between ``inside'' and ``outside'', ``left'' and ``right'', as global properties. Orientation becomes locally meaningful but globally incoherent. This has consequences, developed in Section 7, for the nature of separability and hence for the interpretation of quantum correlation.

\textbf{Pre-metric} means that the regime precedes the introduction of distance, coordinates, or metric relations. It is not a space in the technical sense. It is a domain of structural constraint prior to geometry.

\textbf{Minimal invariant} means that there is some structural identity that persists through the pre-metric regime, something that is preserved under admissible transformations, without presupposing metric structure for its definition.

\textbf{Grounded in} means that the invariant constitutes the foundational structural element from which the closure and the projection are derived. It is not a given background but the minimal condition for there being anything to project. The formal constraints governing projection residue and lag, including quantitative bounds and candidate instantiations, are specified in the Annex (A.3b, A.5b, A.10).

\section{The Three-Part Schema}
The framework is organised around three conceptual elements: the Invariant ($\mathcal{I}$), the Closure ($\mathcal{C}$), and the Projection ($\mathcal{P}$). These do not constitute a completed physical theory. They define a minimal structural schema from which the consequences of Section 4 onwards are derived.

\subsection{The Invariant (\texorpdfstring{$\mathcal{I}$}{I})}
The invariant is the minimal structural element of the framework. It is defined by three negative conditions and one positive one.

Negatively: the invariant has no spatial extension, no metric, and no internal multiplicity. It does not presuppose a background against which it could be located, a distance by which it could be measured, or a collection of which it could be a member.

Positively: the invariant persists. It is the minimal condition for distinguishability (not distinguishability from other things, as multiplicity is not yet available, but the structural fact of there being something rather than nothing) to which subsequent operations apply.

As a heuristic (and we stress that it is heuristic) the invariant may be modelled as a topological loop: a closed path that encodes persistence without requiring coordinates. The loop does not assert that the pre-metric domain is literally one-dimensional. It asserts that closure without metric extension is the minimal structure in play. A loop is the simplest mathematical object that captures this: it is continuous, self-returning, and requires no embedding space to be defined.

The framework's strong claim, discussed further in Section 8, is that there is \textit{one} invariant. Apparent multiplicity (the $\approx 10^{80}$ particles that constitute observable matter) is a consequence of projection, not a feature of the pre-metric domain.

\subsection{The Closure (\texorpdfstring{$\mathcal{C}$}{C})}
The closure introduces the first structural constraint on the invariant. It is the operation by which the invariant generates relational structure without yet generating metric structure.

Formally, the closure can be modelled as a non-orientable mapping of the following character. Take the invariant loop and identify points under a parity-inversion relation. The result is a minimal non-orientable unit: a structure on which orientation is locally definable but globally inconsistent. The mathematical object closest in spirit to this construction is the Möbius strip or, in its closed surface version, the Klein bottle or the real projective plane ($\mathbb{R}P^2$).

The philosophical significance of the closure is this: on a non-orientable surface, two points that appear, under local examination, to be on ``opposite sides'' are in fact connected by a single continuous path. The appearance of opposition is a local artefact. Globally, they are the same structural location traversed from different directions. This observation is the foundation of the framework's treatment of quantum entanglement.

A further structural feature of the closure is its minimal traversal order. The closure need not be parameterised as a metric loop. It may instead be defined by a minimal nontrivial relation $r$ such that $r \neq I$ and $r^2 = I$. This defines a duplex structure: two structural positions related by $r$, where identity is restored only after both positions are traversed under any admissible parameterisation.

\subsection{The Projection (\texorpdfstring{$\mathcal{P}$}{P})}
The projection is not a single monolithic operation but a sequence of layer-specific compressions across dimensional emergence. Different stages preserve different structural features, and the closure compression relevant to observable metric structure occurs specifically at the stage where a pre-metric closure layer is rendered into three-dimensional spatial rotation structure. The staged projection is indexed as $P_{01}, P_{12}, P_{23}$, where $P_{23}$ maps the pre-metric closure layer $S_2$ into the metricated observable manifold $S_3$. The architecture as a whole maps from the pre-metric domain to the three-dimensional metric record:
\begin{equation}
    \mathcal{P}: \mathcal{C} \rightarrow M^3
\end{equation}
where $M^3$ denotes a three-dimensional orientable metricated manifold, the structure we recognise as physical space.

This mapping has three defining features. First, it is surjective but not injective. Second, it is asymmetric. Third, it introduces separability. In $\mathcal{C}$, points that appear opposed are structurally continuous. In $M^3$, after projection, they are separated by a metric distance. Separability is a product of the projection, not a pre-existing feature of the world.

\subsection{Closure Compression and Spinorial Behaviour}
The following result is presented in the conditional register established in the preface. It does not derive spinorial structure from first principles. It shows that if the closure in the pre-metric layer $S_2$ carries a minimal nontrivial relation of order two, then the staged projection architecture developed above produces, as a necessary consequence, a doubled identity cycle in the observable rotational manifold.

The pre-metric closure layer $S_2$ is defined as the closure-complete state layer in which transformation paths exist and identity is defined by structural return, but metric distance, angular coordinate, and classical spatial orientation are not yet fixed. $S_2$ carries the duplex closure relation $r$ introduced in Section 3.2: two structural positions related by $r$, with $r \neq I$ and $r^2 = I$. No traversal parameter $\theta$ lives in $S_2$; the duplex structure is an unparameterised relation over the closure. Any consistent parameterisation of traversal, introduced only when projection renders $S_2$ into the observable manifold $S_3$, will find that two passes are required to return to the starting structural position.

The projection $P_{23}$ maps $S_2$ into $S_3$, where $S_3$ encodes rotation via an $SO(3)$-type observable manifold with intrinsic $2\pi$ periodicity. Four conditions govern the behaviour of $P_{23}$. First, it is surjective and non-injective: closure positions that are structurally distinct in $S_2$ may map to the same observable position in $S_3$. Second, it is closure-minimal: no strictly coarser identification of $S_2$ positions would yield the same observable manifold. Third, $S_3$ is quotient-sufficient at this stage: the observable manifold $S_3$ is the complete observational record, and no additional distinction from $S_2$ survives into the observable domain except through $P_{23}$. Fourth, the projection is cycle-consistent: if identity in $S_2$ requires traversal of the full duplex orbit, the induced observable cycle in $S_3$ cannot count as identity return unless it lifts to a closed identity cycle upstairs.

Under these conditions, the two positions of the duplex closure related by $r$ in $S_2$ cannot be distinguished in $S_3$, because $S_3$ encodes rotation only modulo $2\pi$ and has no representational capacity for their distinction. Projection must therefore identify them. The induced equivalence class is binary, the residual transformation is involutive, and a single observable $2\pi$ cycle corresponds only to passage between the two members of the duplex rather than to full structural return. Identity closure therefore requires two complete observable cycles.

In the observable domain $S_3$, this manifests as spin-\nicefrac{1}{2} behaviour: a $2\pi$ rotation produces a lawful state transformation that is not identity return, a residual sign inversion persists, and a $4\pi$ rotation restores structural identity. The standard spinor transformation law is retained without modification. The projection account adds only the interpretation: the observable $2\pi$ rotation is the projected trace of a pre-metric closure whose minimal identity cycle is double the observable cycle, because the two structural positions of the duplex are identified under projection while remaining distinct in $S_2$. Projection compresses the closure; it does not generate the doubling. The doubling is latent in $S_2$.

The unresolved question is upstream: what fixes the order of the closure relation in $S_2$. The staged projection architecture explains why an order-2 closure in $S_2$ produces a $4\pi$ identity cycle in $S_3$. It does not explain why the closure takes order $2$ rather than some other value. That question is addressed formally in Appendix A.8 and Appendix A.9.

\section{The Fossil Record: 3D Space as the Past}
In the projection framework, the relationship between space and time is inverted. Three-dimensional metricated space is not the arena of events. It is the record of a process. It is the \textit{fossil}.

When the non-orientable closure projects into $M^3$, it leaves a fixed, metricated, orientable structure behind. That structure is the past. It has already been resolved. It cannot be un-projected. The metric, the coordinates, the separable entities that appear within $M^3$ are all artefacts of a projection event that is, in a precise sense, over.

The ``present'' in this picture is not a location in time. It is the edge of the projection process. The ``future'' is the as-yet-unprojected structure of $\mathcal{C}$.

\section{Time as Projection Asymmetry}
Time is not a dimension. Time is not an entropy gradient \citep{penrose1989, carroll2010}. Time is the asymmetry of the projection mapping itself.

Because $\mathcal{P}$ is surjective but not injective, multiple pre-image configurations in $\mathcal{C}$ map to single stabilised outcomes in $M^3$. This mapping has an inherent directionality: it runs from $\mathcal{C}$ to $M^3$. It cannot run in reverse, because $M^3$ lacks the structural richness of $\mathcal{C}$ required to recover the pre-image. Formally, this irreversibility corresponds to the non-trivial kernel of the projection: multiple pre-image configurations mapped to a single outcome cannot be uniquely recovered. The ordering imposed by this irreversible asymmetric mapping is time. Irreversibility is built into the architecture.

\section{Matter as Stabilised Residue}
Matter is not primitive. Matter is the stable residue of projection.

When $\mathcal{P}$ maps $\mathcal{C}$ to $M^3$, the pre-image in $\mathcal{C}$ has structural variation: regions of tighter constraint, of more complex closure. When these project, they produce regions of $M^3$ that are more rigidly structured than their surroundings. These regions are what we observe as matter. Empty space is not nothing. It is a region of the fossil where the projection has left a lighter imprint.

The $\beta$-bound offers a precise way to sharpen this account. Where the projection gate and the observable structure of $M^3$ fail to commute, the resulting residue is quantitatively constrained rather than arbitrary; its magnitude is bounded by the commutator norm $\|[F,E]\|$ via the $\beta$-bound, and this bounded residue is what we call matter (see Annex A.3b).

\section{Entanglement Dissolved: Quantum Correlation as Pre-Separable Unity}
The projection hypothesis dissolves the apparent paradox of quantum entanglement \citep{bell1964, aspect1982} by locating its source in a false presupposition: that spatial separation is a fundamental feature of the world.

Two regions of $M^3$ that are metrically separated were not necessarily distinct in $\mathcal{C}$. In the closure $\mathcal{C}$, configurations that project to well-separated regions of $M^3$ may have been structurally continuous. Their correlation in $M^3$ is a remnant of pre-projection unity.

\section{The Single Invariant and the Moiré of Multiplicity}
This extends the one-electron universe hypothesis associated with Wheeler, grounding it in topology rather than worldline geometry \citep{kragh1990}.

\section{Gravity as Metric Tension}
If matter is stabilised projection residue, then concentrations of matter are regions where the projection from $\mathcal{C}$ has been most active and most dense. The structural gradient between the crystallised fossil record at the matter concentration and the lighter fossil record in the surrounding region is what we observe as gravity.

\section{Black Holes as Projection Saturation}
A black hole is a region in which the projection reaches its limit. As matter density increases, the projection encounters a structural constraint. The interior of a black hole is a region where the projection has saturated, where the mapping $\mathcal{P}$ can no longer produce separable, metricated fossil content. The interior no longer belongs to the domain over which the projection gate can act.

Information is not destroyed at the horizon \citep{hawking1975}. It is withdrawn from the domain of projection. The entropy is surface entropy \citep{bekenstein1973} because the interior is outside the admissible algebra.

\section{Dark Matter as Projection Lag}
Dark matter is not an undetected particle species \citep{rubin1980, zwicky1933}. It is the gravitational signature of incomplete projection.

Different regions of $\mathcal{C}$ project at different rates. This variation can be formalised as a lag field $\lambda$ defined over the projected manifold $M^3$, encoding local projection latency. Regions of non-zero lag correspond to configurations that have begun to influence the metric structure but have not fully resolved into the admissible algebra. The resulting effect is gravitational without electromagnetic signature, as full projection is a condition of separability. A precise definition of $\lambda$ and its linear response regime is given in Annex A.5b.

\section{The Open Problem: Cyclic Constraint and the Evolution of Projection}
Does the projection preserve the invariant strictly, or does each cycle of projection impose constraints on $\mathcal{C}$ that shape subsequent emergence? If the latter (the constraint-imposition reading), the universe ``learns'' its structure, and the laws of physics are the accumulated consequence of prior projection cycles. This is the central open problem of the framework.

\section{Summary of the Conditional Consequences}

\begin{table}[h!]
	\caption{Structural Comparison: Prevailing Physics vs. Projection Framework}
	\centering
	\begin{tabular}{p{3cm} p{5.5cm} p{6.5cm}}
		\toprule
		\textbf{Phenomenon} & \textbf{Prevailing Physical Treatment} & \textbf{Projection Account} \\
		\midrule
		Time's arrow & Entropy gradient / low initial entropy & Asymmetry of non-injective projection \\
		Matter & Ontologically primitive particles/fields & Stabilised residue of projection density \\
		Empty space & Absent matter / vacuum state & Light projection fossil \\
		Quantum entanglement & Non-local correlation / paradox & Remnant of pre-projection structural continuity \\
		Multiplicity & Fundamental plurality of entities & Moiré effect of single invariant under projection \\
		Gravity & Spacetime curvature / force & Metric tension at projection density gradient \\
		Black holes & Information paradox / extreme curvature & Projection saturation (loss of admissible representation) \\
		Dark matter & Unknown particle species & Projection lag (incomplete resolution of $\mathcal{C}$ into $M^3$) (formalised via lag field $\lambda$, see A.5b) \\
		The present & Point on timeline & Active frontier of projection process \\
		The past & Elapsed events & Fixed fossil record: metricated, irreversible \\
		The future & Awaiting determination & Unprojected pre-metric domain \\
		\bottomrule
	\end{tabular}
	\label{tab:comparison}
\end{table}

\section{Conclusion}
This paper has proposed that if three-dimensional spacetime arises through asymmetric projection from a non-orientable pre-metric regime grounded in a minimal invariant, a coherent and internally constrained set of consequences follows. The framework does not claim to have solved these problems; it reorganises them around a single structural hypothesis whose consequences are neither trivial nor optional once granted.

\appendix
\section{Candidate Formal Objects and Research Obligations}

This annex presents candidate mathematical structures corresponding to each element of the projection framework. These structures are presented as directional proposals rather than completed derivations. The research obligations identified are the proof requirements that would be needed to elevate the conditional claims of the main paper to formal results. The annex does not claim that these formal objects exist in the required form; establishing that they do is the work to be done.

On formal pathways. The framework already has nearby formal relatives that sharpen its research obligations. The gate/measurement formalism developed in \citep{sendall2026} provides an operational language for the projection: the coherence witness $W(\rho,F)=\|F\rho(I-F)\|_1$ measures cross-boundary coherence of a state relative to the gate; the record fidelity gap $\Delta_T(\rho_F,R)$ measures how much a readout's expectation value changes under symmetrisation; and the $\beta$-bound controls the structured distortion that gate/readout mismatch produces. These are not yet deployed as results in the main paper; they are identified here as the most direct route by which the conditional claims of the main text could be given formal standing.

\subsection{The Invariant: Candidate Formal Object}
Let $\mathbf{C}$ be a category. The invariant $\mathcal{I}$ is an object of $\mathbf{C}$ provisionally modelled as having no nontrivial endomorphisms.
\textit{Research obligation:} Characterise the category $\mathbf{C}$ and establish the admissible transformation class that preserves $\mathcal{I}$.

\subsection{The Closure: Candidate Formal Object}
The closure is modelled as a non-orientable quotient construction producing a minimal non-orientable unit:
\begin{equation}
    U = (\gamma \times [0,1]) / \sim
\end{equation}
where $(x, 0) \sim (\tau(x), 1)$ for an orientation-reversing involution $\tau$.
\textit{Research obligation:} Characterise the space of admissible closures and show that self-intersection to embed in $\mathbb{R}^3$ corresponds precisely to temporal ordering.

\subsection{The Projection and Gate Formalism}
The projection corresponds to a conditional expectation $F_{adm}: \mathcal{A}(\Sigma) \rightarrow \mathcal{A}_{adm}$. The full algebra $\mathcal{A}(\Sigma)$ encodes the pre-metric domain; the accessible algebra $\mathcal{A}_{adm}$ encodes the fossil record $M^3$.
The dimensional discrepancy tensor is defined as:
\begin{equation}
    D_{\mu\nu} := G_{\mu\nu}[g_{3D}] - 8\pi G \bar{T}_{\mu\nu}
\end{equation}
where $D_{\mu\nu}$ captures the gravitational effect of incompletely projected lag regions.
\textit{Research obligation:} Derive the conservation $\nabla^\mu D_{\mu\nu} = 0$ from the architecture of $\mathcal{P}$, and link it formally to the $\beta$-bound.

\subsection*{A.3b The \texorpdfstring{$\beta$}{beta}-bound as a Structural Constraint}
\addcontentsline{toc}{subsection}{A.3b The \texorpdfstring{$\beta$}{beta}-bound as a Structural Constraint}

Let $F$ denote the projection gate and $E_{\mu\nu}$ a local observable in the accessible algebra. The deviation between projected and fully resolved expectation values satisfies:
\begin{equation}
    |\Delta p_F(E_{\mu\nu})| \leq 2 \sqrt{\frac{1-s}{s}} \|[F, E_{\mu\nu}]\|
\end{equation}
This bound establishes that projection-induced discrepancy is not arbitrary. Its magnitude is controlled by the failure of the projection gate and observable structure to commute.

The dimensional discrepancy tensor $D_{\mu\nu}$ defined in (3) is therefore not freely specifiable. Its magnitude is constrained by the algebraic structure of the gate.

Interpretively, where commutation fails, the resulting deviation is structured residue rather than noise. The $\beta$-bound and the dimensional discrepancy tensor $D_{\mu\nu}$ therefore form a closed formal loop: non-commutation at the level of the projection gate produces bounded residue, and that residue appears as curvature in excess of baryonic sources.

\textit{Research obligation:} determine whether gravitational anomalies can be classified in terms of commutator norms within a concrete gate model.

\subsection{Black Holes: Formal Obligations}
The event horizon is the surface at which the interior algebra passes outside $\mathcal{A}_{adm}$, the boundary beyond which the gate $F_{adm}$ can no longer act. The bridge lemma for bipartite systems, $\mathrm{Tr}_A[(F_A^* \otimes \mathrm{id})(\rho_{AB})] = \rho_B$, ensures this exclusion is consistent with no-signalling.

\stepcounter{subsection}
\subsection*{A.5b The Lag Field and Dark Matter}
\addcontentsline{toc}{subsection}{A.5b The Lag Field and Dark Matter}

Let $\lambda$ be a real-valued function on the admissible algebra $\mathcal{A}_{adm}$, or equivalently a scalar field on the projected manifold $M^3$, encoding local projection latency.

For a family of perturbations $p$ of the pre-metric domain, the induced change in primitive geometric quantities satisfies a bounded linear response condition:
\begin{equation}
    \sup_{\gamma} |\Delta \ell(\gamma; p)| < \varepsilon_{\text{resp}}
\end{equation}
for some finite response radius $\varepsilon_{\text{resp}}$.

Within this regime, lag behaves linearly and enters the effective metric structure as a time-change of geodesic flow. Outside this regime, nonlinear accumulation of unresolved structure is expected.

Regions of high $\lambda$ correspond to incomplete projection: configurations that influence the metric structure without full admissible representation.

In the weak-field limit, the gradient $\nabla \lambda$ produces observable effects consistent with baryonic scaling relations. One candidate refinement is to relate $\lambda(x)$ to the dimensionality of unresolved pre-image structure, for example via the effective dimension of the kernel of the projection map at the corresponding location.

\textit{Research obligation:} derive observable consequences of $\nabla \lambda$ beyond the linear regime and determine empirical distinguishability from particle dark matter models.

\subsection{The Open Problem: Formal Framing}
The open problem asks if $F_{adm}$ is strictly idempotent ($F^2 = F$) or quasi-idempotent ($F^2 \approx F + \epsilon$). If $\epsilon \neq 0$, the gate itself carries a history.

\subsection{Closure Compression Result: Formal Statement}
Let $P_{23}: S_2 \rightarrow S_3$ be a surjective, non-injective projection. If $S_2$ carries a duplex closure relation $r \neq I$ and $r^2 = I$, the induced equivalence classes under $P_{23}$ are binary, forcing spin-\nicefrac{1}{2} observable behaviour. The justification for the order-2 assumption is developed in A.8 and A.9.

\subsection{Upstream Obligation: The Order of Closure in \texorpdfstring{$S_2$}{S2}}
Why does $S_2$ carry an order-2 relation? If the only structural distinction available in a pre-metric layer without metric is the binary distinction between a state and its structural complement, then the minimal nontrivial closure orbit has exactly two members.

\subsection{Pre-Metric Closure Order Result (The Logic of Squeeze)}

\begin{description}
    \item[The ``At Least'' Constraint:] $S_2$ must introduce at least one non-identity relation ($r \neq I$).
    \item[The ``At Most'' Constraint:] Without metric or indexing structure, $S_2$ cannot individuate more than one non-identity position.
    \item[Forced Involution:] A two-member orbit $\{s, r(s)\}$ forces $r(r(s)) = s$, thus $r^2 = I$. Spin-\nicefrac{1}{2} is not assumed; it is the minimal non-degenerate structural solution for an emergence stage constrained by relational distinguishability.
\end{description}

\subsection*{A.10 A Worked Example: Lag from Prime Geodesic Structure}
\addcontentsline{toc}{subsection}{A.10 A Worked Example: Lag from Prime Geodesic Structure}

To demonstrate that the lag field admits concrete instantiation, consider a compact hyperbolic surface:
\begin{equation}
    X = \Gamma \backslash \mathbb{H}
\end{equation}
where $\Gamma$ is a discrete subgroup of $\mathrm{PSL}(2,\mathbb{R})$.

Let $p$ denote primitive closed geodesics on $X$, with lengths $\ell(p)$. These define a discrete lag spectrum:
\begin{equation}
    T_\omega = \{\ell(p)\}
\end{equation}

Introduce a $\mathbb{Z}_2$ character $\chi: \Gamma \rightarrow \{\pm 1\}$.

The Green kernel of the twisted Laplacian $\Delta_\chi + m^2$ admits an expansion:
\begin{equation}
    G_\chi(x,y) = \sum_{\gamma \in \Gamma} \chi(\gamma) \, g_m(d(x, \gamma y))
\end{equation}
which reorganises into a sum over primitive geodesics:
\begin{equation}
    G_\chi(x,y) \sim \sum_{p} \sum_{k=1}^\infty \chi(p)^k \, g_m(k\ell(p))
\end{equation}

In this construction, primitive geodesic lengths act as lag parameters, and the twist encodes projection asymmetry.

This provides a concrete example in which incomplete projection produces cumulative, computable effects at larger scales.

\textit{Research obligation:} determine whether physically relevant lag fields can be derived from spectral data of admissible structures and whether such constructions admit empirical discrimination.

\bibliographystyle{unsrtnat}

\begin{thebibliography}{9}

\bibitem[Bekenstein(1973)]{bekenstein1973}
Bekenstein, J. D. (1973). Black holes and entropy. \textit{Physical Review D}, 7(8), 2333.

\bibitem[Kragh(1990)]{kragh1990}
Kragh, H. (1990). \textit{Dirac: A Scientific Biography}. Cambridge University Press.

\bibitem[Hawking(1975)]{hawking1975}
Hawking, S. W. (1975). Particle creation by black holes. \textit{Communications in Mathematical Physics}, 43(3), 199-220.

\bibitem['t Hooft(1993)]{thooft1993}
't Hooft, G. (1993). Dimensional reduction in quantum gravity. \textit{arXiv preprint gr-qc/9310026}.

\bibitem[Susskind(1995)]{susskind1995}
Susskind, L. (1995). The world as a hologram. \textit{Journal of Mathematical Physics}, 36(11), 6377-6396.

\bibitem[Jacobson(1995)]{jacobson1995}
Jacobson, T. (1995). Thermodynamics of spacetime: the Einstein equation of state. \textit{Physical Review Letters}, 75(7), 1260.

\bibitem[Verlinde(2011)]{verlinde2011}
Verlinde, E. (2011). On the origin of gravity and the laws of Newton. \textit{Journal of High Energy Physics}, 2011(4), 1-27.

\bibitem[Padmanabhan(2010)]{padmanabhan2010}
Padmanabhan, T. (2010). Thermodynamical aspects of gravity: new insights. \textit{Reports on Progress in Physics}, 73(4), 046901.

\bibitem[Penrose(1989)]{penrose1989}
Penrose, R. (1989). \textit{The Emperor's New Mind}. Oxford University Press.

\bibitem[Carroll(2010)]{carroll2010}
Carroll, S. (2010). \textit{From Eternity to Here: The Quest for the Ultimate Theory of Time}. Dutton.

\bibitem[Bell(1964)]{bell1964}
Bell, J. S. (1964). On the Einstein Podolsky Rosen paradox. \textit{Physics Physique Fizika}, 1(3), 195.

\bibitem[Aspect et al.(1982)]{aspect1982}
Aspect, A., Dalibard, J., \& Roger, G. (1982). Experimental test of Bell's inequalities using time-varying analyzers. \textit{Physical Review Letters}, 49(25), 1804.

\bibitem[Rubin et al.(1980)]{rubin1980}
Rubin, V. C., Ford, W. K. J., \& Thonnard, N. (1980). Rotational properties of 21 SC galaxies. \textit{Astrophysical Journal}, 238, 471.

\bibitem[Zwicky(1933)]{zwicky1933}
Zwicky, F. (1933). Die Rotverschiebung von extragalaktischen Nebeln. \textit{Helvetica Physica Acta}, 6, 110-127.

\bibitem[Sendall(2026)]{sendall2026}
Sendall, J. (2026). The Beta-Bound: Drift constraints for Gated Quantum Probabilities. \textit{arXiv preprint arXiv:2601.22188}. \url{https://arxiv.org/abs/2601.22188}

\end{thebibliography}

\end{document}